\documentclass[
12pt,
pra,
a4paper,
amsmath,
floatfix,
notitlepage,
longbibliography
]{revtex4-1}

\usepackage{graphicx,amssymb,amsmath,amsfonts}
\usepackage{epsfig}
\usepackage{xcolor}
\usepackage[mathscr]{eucal}
\usepackage{comment}
\usepackage{bm}

\usepackage{xr}
\let \ShowFixme = 0
\let \Date = 1
 \usepackage{grffile}

\usepackage{xcolor}

\usepackage[draft,inline,nomargin]{fixme}
\fxsetup{theme=color,mode=multiuser}
\FXRegisterAuthor{sd}{asd}{\color{red}SD}
\FXRegisterAuthor{ov}{aov}{\color{blue}OV}
\FXRegisterAuthor{sk}{ask}{\color{orange}SK}
\FXRegisterAuthor{all}{aall}{\color{blue}\underline{COMMENT}}

\newlength{\figwidth}

\newif\ifOnecolumn
  \ifx\Onecolumn\undefined
  \Onecolumntrue
  
\else
  \Onecolumnfalse
\fi

\newif\ifTwocolumn
  \ifOnecolumn
   \Twocolumnfalse
\else
  \Twocolumntrue
\fi

\usepackage{xspace}
\newcommand{\latin}[1]{{\it #1}}
\newcommand{\ie}{\latin{i.e.}\@\xspace}
\newcommand{\eg}{\latin{e.g.}\@\xspace}
\newcommand{\cf}{\latin{cf.}\@\xspace}
\newcommand{\etc}{\latin{etc.}\@\xspace}

\def\ba{\begin{align}}
\def\ea{\end{align}}
\def\la{\langle}
\def\ra{\rangle}

\newcommand{\ftitle}[1]{\textit{#1}}
\newcommand{\fig}[1]{Fig.~\ref{#1}}

\newcommand{\figss}[1]{Figs.~\ref{#1}}

\newcommand{\sfig}[1]{Fig.~\ref{si:#1}}
\newcommand{\Fig}[1]{Figure~\ref{#1}}

\newcommand{\myref}[1]{Ref.~\cite{#1}}
\newcommand{\myrefs}[1]{Refs.~\cite{#1}}

\newcommand{\eq}[1]{Eq.~(\ref{#1})}

\renewcommand{\vr}{\boldsymbol{r}}

\renewcommand{\H}{\mathcal{H}}

\begin{document}

\title{Bridging transitions and capillary forces for colloids in a slit}

\author{Oleg A. Vasilyev}
\author{Marcel Labb\'{e}-Laurent}
\author{S. Dietrich}

\affiliation{Max-Planck-Institut f{\"u}r Intelligente Systeme, Heisenbergstra{\ss}e~3, D-70569 Stuttgart, Germany}
\affiliation{IV. Institut f\"ur Theoretische Physik, Universit{\"a}t Stuttgart,  Pfaffenwaldring 57, D-70569 Stuttgart, Germany}

\author{Svyatoslav Kondrat}
%\email{s.kondrat@fz-juelich.de}
%\affiliation{Institute for Computational Physics, University of Stuttgart, Allmandring 3, 70569 Stuttgart, Germany}
\affiliation{Department of Complex Systems, Institute of Physical Chemistry, Polish Academy of Sciences, Kasprzaka 44/52, 01-224 Warsaw, Poland}

\begin{abstract}

    Capillary bridges can form between colloids immersed in a two phase fluid, \eg, in a binary liquid mixture, if the surface of the colloids prefers the species other than the one favored in the bulk liquid. Here, we study the formation of liquid bridges induced by confining colloids to a slit, with the slit walls having a preference opposite to the one of the colloid surface. Using mean field theory, we show that there is a line of first-order phase transitions between the bridge and the no-bridge states, which ends at a critical point. By decreasing the slit width, this critical point is shifted towards smaller separations between the colloids. However, at very small separations, and far from criticality, we observe only a minor influence of the slit width on the location of the transition. Monte Carlo simulations of the Ising model, which mimics incompressible binary liquid mixtures, confirm the occurrence of the bridging transitions, as manifested by the appearance of bistable regions where both the bridge and the no-bridge configurations are (meta)stable. Interestingly, we find no bistability in the case of small colloids, but we observe a sharpening of the transition when the colloid size increases. In addition, we demonstrate that the capillary force acting between the colloids can depend sensitively on the slit width, and varies drastically with temperature, thus achieving strengths orders of magnitude higher than at criticality of the fluid.

\end{abstract}

\maketitle

\section{Introduction}

Restricted geometries and confinement play an important role in physics and lead to a myriad of new phenomena which do not occur in bulk systems. Examples are numerous and include unusual drying~\cite{singh:nature:06:ConfWater} and freezing~\cite{koga:nature:01:ConfWaterFreezing, han:nphys:10:ConfWater} of confined water, quantum \cite{Casimir:1948} and critical \cite{Fisher-et:1978, Gambassi:2009, Maciolek2018} Casimir effects, voltage-induced phase transitions in confined ionic liquids \cite{kondrat:jpcm:11, lee16a}, long-ranged repulsion between confined dimers~\cite{tkachenko:prl:17:LongRangeRepulsion}, \etc A particularly intensively studied system is a simple fluid, or a binary liquid mixture, in restricted geometries. It is well known that, if a two-phase fluid is confined to a slit (without colloids), with both slit walls favoring the same phases, the system undergos capillary condensation, in that the slit becomes filled by the preferred phase \cite{Evans:1990, Gelb-et:1999, Binder-et:2008}. A similar effect can be observed between any kind of bodies, \eg, colloids, having the same preference for a fluid phase, but opposite to what is preferred in the bulk liquid. In this case, these bodies may become connected via capillary bridges, which consist of the fluid phase in accordance with their surface preference. Such bridges lead to strong capillary forces, which are crucial for studies of powders, soils, granular materials, friction, stiction \etc \cite{Butt2009}, and can be used to establish colloidal networks and other microstructures \cite{Oliver2000, koos:curropcoll:14:CollSuspensions, Cheng2012}.

Capillary bridging has been extensively investigated theoretically \cite{Haines1925, Fisher1926, Dobbs1992, Dobbs1992a, Bauer-et:2000, Andrienko2004, Archer-et:2005, Grof2008, Hopkins-et:2009, Dutka-et:2010, Cheng2012, Okamoto-et:2013, Saavedra2014, malijevsky15, Farmer2015, Labbe-Laurent:17:jcp, Chacko2017, vasilyev:18:sm} and experimentally \cite{Haines1925, Mason1965, Beysens-et:1985, Willett2000, Oliver2000, Butt2009, Goegelein-et:2010, Ataei2017}. For instance, already in the mid-1920s, \citeauthor{Haines1925} \cite{Haines1925} and \citeauthor{Fisher1926} \cite{Fisher1926} realized the importance of capillary forces in soils and provided analytical expressions for the force. \citeauthor{Mason1965} \cite{Mason1965} measured the force between two colloids and a colloid and a wall as a function of separation for different bridge volumes. \citeauthor{Beysens-et:1985} \cite{Beysens-et:1985} observed the aggregation of silica colloids in a mixture of water and lutidine; the aggregation was likely due to the capillary forces emerging in the water-lutidine coexistence region.

The formation and breaking of capillary bridges can proceed continuously or via a phase transition \cite{Dobbs1992, Dobbs1992a, malijevsky15, Chacko2017, Labbe-Laurent:17:jcp, Nowakowski2014, Nowakowski2016}. For instance, \citeauthor{Dobbs1992} \cite{Dobbs1992, Dobbs1992a} used an effective interface model and found a first-order bridging transition for two spherical colloids. \citeauthor{malijevsky15} \cite{malijevsky15} studied the bridge formation for spherical and cylindrical particles by using macroscopic and microscopic approaches. They found that the transitions are strongly first order for cylinders but can be first or second order for spheres; interestingly, for small spheres the transformation between the bridge and no-bridge states was continuous (\ie, without a transition). \citeauthor{Labbe-Laurent:17:jcp} \cite{Labbe-Laurent:17:jcp} applied a mean-field theory for cylindrical particles and also found first-order bridging transitions.  \citeauthor{Chacko2017} \cite{Chacko2017} used classical density functional theory to study solvent-mediated interactions between two blocks immersed in a simple one-component fluid. They showed, in particular, that the force acting between two solvophobic blocks exhibits a jump upon varying the block-block separation when the number density between the blocks changes from gas-like (bridge) to liquid-like (no bridge), indicative of a first-order bridging transition. Remarkably, \citeauthor{Nowakowski2014} \cite{Nowakowski2014, Nowakowski2016} solved analytically the two-dimensional Ising model in a strip and studied the bridge formation between chemically inhomogeneous walls. Although this formation process was found to be continuous, \ie, without a transition, in certain limiting cases it can turn into a first-order transition \cite{nowakowski:personal}.

An efficient way to study capillary bridging between colloids is to confine them into a slit, with the slit walls exhibiting preferences for a fluid phase which compete with those of the colloid surfaces. Such a setup has been considered in \myref{vasilyev:18:sm}, which focused on non-additivity of colloid-colloid interactions if a binary liquid mixture is in the mixed phase. This study also showed that capillary bridges can form between the colloids in the demixed, two-phase region. Here, our goal is to scrutinize the formation and breaking of \emph{such} bridges. We are particularly interested in bridging \emph{transitions} and colloid-colloid interactions, and their dependences on various parameters, such as temperature and the slit width. To this end, we use mean field theory in order to provide a global bridging phase diagram and to calculate the capillary forces acting between the colloids. In order to asses the role of fluctuations in this system, we perform Monte Carlo simulations for the same setup by using the Ising model, which mimics an incompressible binary liquid mixture or a simple one-component fluid.

\section{Model and Methods}
\label{sec:methods}

We consider two identical spherical colloids of radius $R$ confined to a slit of width $w$ filled with a two-phase fluid (\cf \fig{fig:mft}(a)). For both mean field theory and Monte Carlo simulations, we assume the colloids to have boundary conditions opposite to those on the slit surfaces, so that they prefer different species of a binary liquid mixture (or opposite values of the spin magnetization in the Ising model). This setup allows for the formation of capillary bridges between the colloids.

\subsection{Mean field theory}
\label{sec:methods:mft}

We describe a two-phase fluid by the standard dimensionless Ginzburg-Landau-Wilson (LGW) Hamiltonian in spatial dimension $d$:
\begin{align}
\label{eq:mft:H}
	\H[\varphi] = \int_V d^d r \left\{\frac{1}{2}(\nabla \varphi)^2 + \frac{\tau}{2} \varphi^2 + \frac{g}{4!}\varphi^4 \right\} ,
\end{align}
where $d^d r=\prod_{i=1}^{d} dx_i$ is the $d$-dimensional volume element, $V$ is the volume accessible to the fluid, and the coupling constant $g>0$ stabilizes $\H$ below the (upper) critical point $T_c$. The order parameter $\varphi(\vr)$ is the difference between the concentrations of the two species for a binary liquid mixture close to demixing (or the magnetization for an Ising magnet, see below). Within mean field theory (MFT) one has $\tau = t/(\sqrt{2}\xi_0^-)^2$ where $t=(T-T_c)/T_c$ is the reduced temperature and $\xi_0^-$ the nonuniversal amplitude of the bulk correlation length $\xi_-=\xi_0^{-}|t|^{-\nu}$ in the ordered phase ($\nu=1/2$ within MFT). Here, we focus solely on the ordered phase, \ie, $\tau < 0$, which describes a two-phase fluid. We note that, on the mean-field level, \eq{eq:mft:H} yields the lowest order of a systematic expansion in terms of $\epsilon = 4-d$ of universal quantities such as critical exponents and scaling functions. Accordingly, the present mean field study can be viewed as the important first step within a general scheme.

We assume all surfaces to belong to the so-called extraordinary (or normal) surface universality class, which corresponds to the limit of an infinitely strong surface field (which expresses the preference of the surface). This implies that the equilibrium concentration profile diverges upon approaching the surface (see, \eg, Refs.~\cite{Binder:1983, Diehl:1986, Mohry-et:2010}). In order to deal with this numerical challenge of minimizing $\H[\varphi]$ under the aforementioned boundary condition, we have used a short-distance expansion (see Refs.~\cite{kondrat:jcp:07, Kondrat-et:2009}) for calculating the values of the order parameter at small distances $\delta$ from the colloidal surface (we have used $\delta = 0.05\times R$ in all calculations); we kept these values fixed in the course of minimization. This approach has proven to be successful in a number of previous studies~\cite{Kondrat-et:2009, Troendle-et:2009, Troendle-et:2010, mohry:softmat:14, LabbeLaurent-et:2014, Labbe-Laurent:17:jcp, vasilyev:18:sm, kondrat:jpcm:18}. We have used the three-dimensional finite element library F3DM~\cite{f3dm} for minimizing numerically the LGW Hamiltonian (\eq{eq:mft:H}). In order to calculate the force acting between colloids, we enclosed the colloids by ellipsoidal surfaces and computed the force by integrating the stress tensor over these surfaces \cite{Kondrat-et:2009}. 

We use the temperature scaling variable $\theta = - (R/\xi_-)^{1/\nu} = t (R/\xi_0^-)^{1/\nu} $ in order to discuss consistently the mean field and Monte Carlo simulation results; note that $\theta = 2 \tau R^2$ within MFT. The slit width $w$ and the colloid-colloid separation $D$ are everywhere expressed in units of the colloid radius $R$.

\subsection{Monte Carlo simulations of the Ising model}

The Ising model mimics an incompressible binary liquid mixture or a simple fluid; its Hamiltonian is given by
\begin{align}
\label{eq:ham_b}
H(\{ \sigma \})= -J \sum \limits_{\la ij \ra }\sigma_{i} \sigma_{j},
\end{align}
where $\sigma_{i}= \pm 1$ is a classical spin at lattice site $i$ so that $\sigma_{i}=\pm 1$ corresponds to occupation of site $i$ by an A particle ($\sigma_i = +1$) or B particle ($\sigma_i = -1$) of a binary liquid mixture and $J$ is the coupling constant. The sum $\la ij \ra$ runs over all neighboring pairs of spins. We studied systems consisting of $N_{x} \times N_{y} \times N_z$ lattice sites, periodic in the lateral $x$ and $y$ directions but of finite width $w=N_z a$ in the $z$ direction, where $a$ is the lattice constant. As within MFT, the system contains two colloids placed a distance $D$ apart, with their centers located in the midplane of the slit (as before, $D$ is the surface-to-surface distance). Within this lattice model, a colloid is defined by drawing a sphere of radius $R$ with its center located at a lattice site and with all spins within the sphere treated as being frozen~\cite{vasilyev:pre:2014:colloidsMC}. For the slit walls and the colloid surfaces, we considered the strong adsorption limit \cite{Binder:1983, Diehl:1986, Mohry-et:2010}, which amounts to fixing the spins on the lattice sites next to the sites of their surface. We also assumed the colloids to have the surface spins oriented opposite to those on the slit walls, in order to facilitate the formation of a liquid bridge between the colloids.

The system size was $N_x = N_y = 160$ lattice sites in the lateral ($x,y$) directions. We checked possible finite size effects by using $N_x = N_y = 260$ (the size in the $z$ direction was fixed by the slit width), for which we have not observed any significant deviations from the results for smaller lattice sizes. The simulations were performed by using $10^6$ hybrid MC steps, with each step consisting of a combination of the Wolff cluster update and single spin Metropolis updates.

The force at separation $D$ was determined via the finite difference $f (D)= - [U(D+1)-U(D-1)]/(2a)$, where the interaction potential $U(D) $ was obtained via integration of the local magnetization over a local field (see \myref{vasilyev2015monte} for details). We note that this method does not work close to a first-order transition due to difficulties in performing the integration.

As within MFT, we introduce the temperature scaling variable $\theta = - (R/\xi_-)^{1/\nu} = t (R/\xi_0^-)^{1/\nu}$, where $t$ is the reduced temperature $t = (T-T_c)/T_c =(\beta_c-\beta)/\beta$ with the inverse dimensionless temperature $\beta = J/(k_B T)$. The critical point is at $\beta_c \approx 0.22165455(3)$ \cite{Deng:03:pre}, the amplitude of the bulk correlation length in the ordered phase is $\xi_0^- =0.2431(1) $ in units of the lattice constant $a$ \cite{Ruge-et:1994}, and the correlation length critical exponent is $\nu \approx 0.63002(10)$ \cite{Hasenbusch-nu}.

%%%%%%%%%%%%%%%%%%%%%%%%%%%%%%%%%%%%
\section{Results and discussions}
%%%%%%%%%%%%%%%%%%%%%%%%%%%%%%%%%%%%

%%%%%%%%%%%%%%%%%%%%%%%%%%%%%%%%%%%%%%%%%%%%%%%%%%%%%%%%%%%%%%%
\subsection{Bridging transitions and mean field phase diagram}

\begin{figure}[t]
    \vspace{-0.6cm}
\begin{center}
\includegraphics[width=\textwidth]{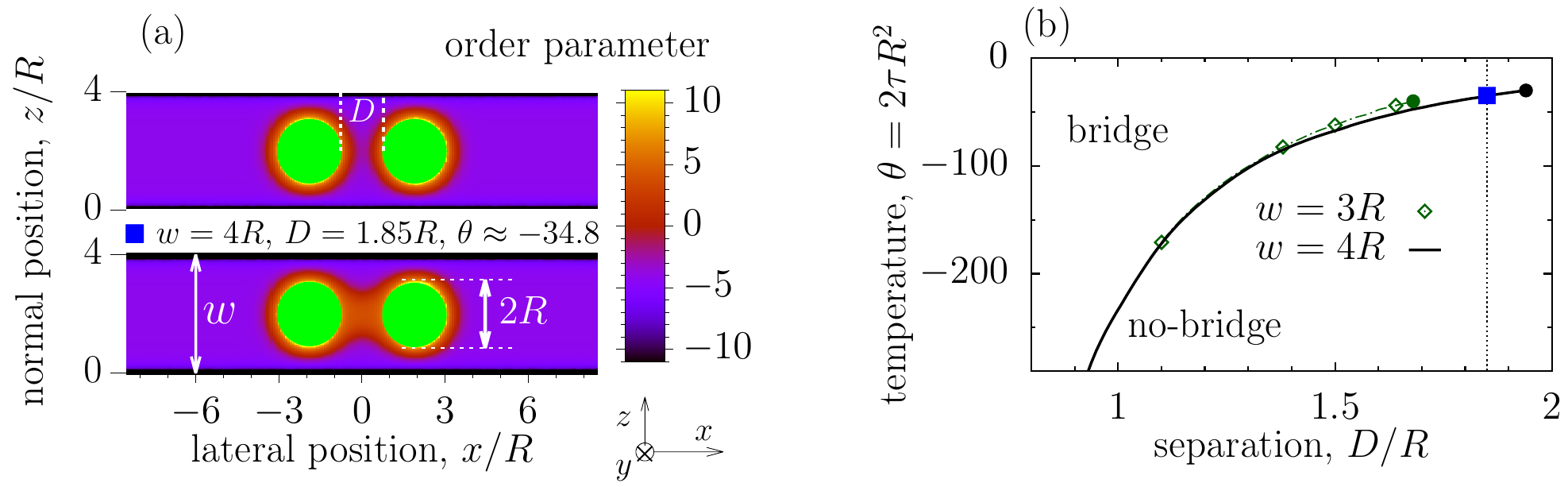}
    \caption{ \ftitle{Bridging transition for confined colloids from mean field theory.} (a) Order parameter distribution for the coexisting bridge and no-bridge configurations of two colloids confined symmetrically to a slit. The boundary conditions on the slit walls are opposite to those on the colloid surfaces. The numerical data correspond to a slit of width $w=4R$ and to a surface-to-surface distance $D=1.85 R$ between the colloids, where $R$ is the colloid radius. The reduced temperature is given by $\theta = 2R^2 \tau \approx -34.8$. The prefactor of the order parameter profile is expressed in terms of the amplitude $\mathcal{A} = \sqrt{3/[g(\xi_0^-)^2]}$  of the bulk order parameter $\varphi_{\mathrm{bulk}} = \mathcal{A} (-t)^{1/2}$ in the ordered phase. (The sidebar provides the magnitude of the local order parameter divided by $\mathcal{A}$.) (b) Bridging phase diagram in the plane spanned by the rescaled temperature $\theta = 2 \tau R^2$ and the colloid-colloid separation $D/R$. The diagram shows the regions of the bridge and no-bridge states separated by a line of first-order transitions. The full circles denote the critical points for $w=3R$ and $w=4R$. The dotted vertical line indicates the value of the colloid separation $D/R=1.85$ used in (a) and in \fig{fig:force}(a). The filled square denotes the value of the colloid separation ($D/R=1.85$) and of the rescaled temperature ($\theta \approx - 34.8$) used in (a), which corresponds to coexistence of the bridge and the no-bridge configuration for $w=4R$. See Supplementary Videos for the formation and the breaking of the bridges as a function of temperature.
    \label{fig:mft}
}
\end{center}
\end{figure}

    The numerical mean field calculations reveal the occurrence of two (meta)stable configurations (\fig{fig:mft}(a)) \cite{vasilyev:18:sm}. In one configuration, a bridge forms between the colloids, which is rich in the species preferred by the colloid surfaces. In the second configuration there is no bridge, but the colloids are surrounded by coronas, the thickness of which is comparable to the correlation length $\xi_-$, and which consist of species favored by the colloid surfaces. In certain parameter ranges, we have found a first-order phase transition between the bridge and the no-bridge states, accompanied by the corresponding metastable branches (\sfig{fig:mft:fe}, and, \cf, \fig{fig:force}(a)). We have computed a global `bridging' phase diagram, which is shown in \fig{fig:mft}(b) in the plane spanned by the reduced temperature and the colloid-colloid separation. The phase diagram consists of a line of first-order transitions, which ends at a critical point (of the bridging transitions) at large separations, and extends to small separations as the temperature is decreased to further negative values of $\theta$, which correspond to the ordered state.

    This structure of the phase diagram (\fig{fig:mft}(b)) is similar to the one obtained for systems without a slit \cite{Okamoto-et:2013, malijevsky15, Labbe-Laurent:17:jcp} (\eg, if capillary bridges are due to an external bulk field, corresponding to an off-critical composition of a binary liquid mixture). In a slit, the phase diagram can additionally be influenced by the slit width. However, at small separations, \ie, $D/R \lesssim 1/5$, this influence appears to be minuscule (see the open diamonds for $w=3R$ and the line for $w=4R$ in \fig{fig:mft}(b)). This is understandable because in this regime the transitions occur at low temperatures (\ie, far from the critical point $\theta =0$ of the fluid), at which the correlation length is small, as compared to the colloid radius and the slit width. Thus the colloids are either surrounded by very narrow coronas, or they are connected by a \emph{narrow} bridge (\sfig{fig:mft:op}). Therefore, in this case, the role of the slit is merely to induce an order parameter far from the colloids to be opposite to the one preferred by the colloids, which facilitates the formation of bridges as such.

    The influence of the slit becomes more pronounced for larger colloid-colloid separations and closer to criticality at $\theta = 0$. In this case, reducing the slit width induces a significant shift of the bridging critical point towards smaller separations. This is the case because the close vicinity of slit walls destabilizes the bridge, turning the transition into a continuous transformation between a bridge-like and the no-bridge state ($D/R \gtrsim 1.7$ for $w=3R$ in \fig{fig:mft}(b)). 

\subsection{Bridging transitions from Monte Carlo (MC) simulations}

\begin{figure}[t]
\begin{center}
\includegraphics[width=\textwidth]{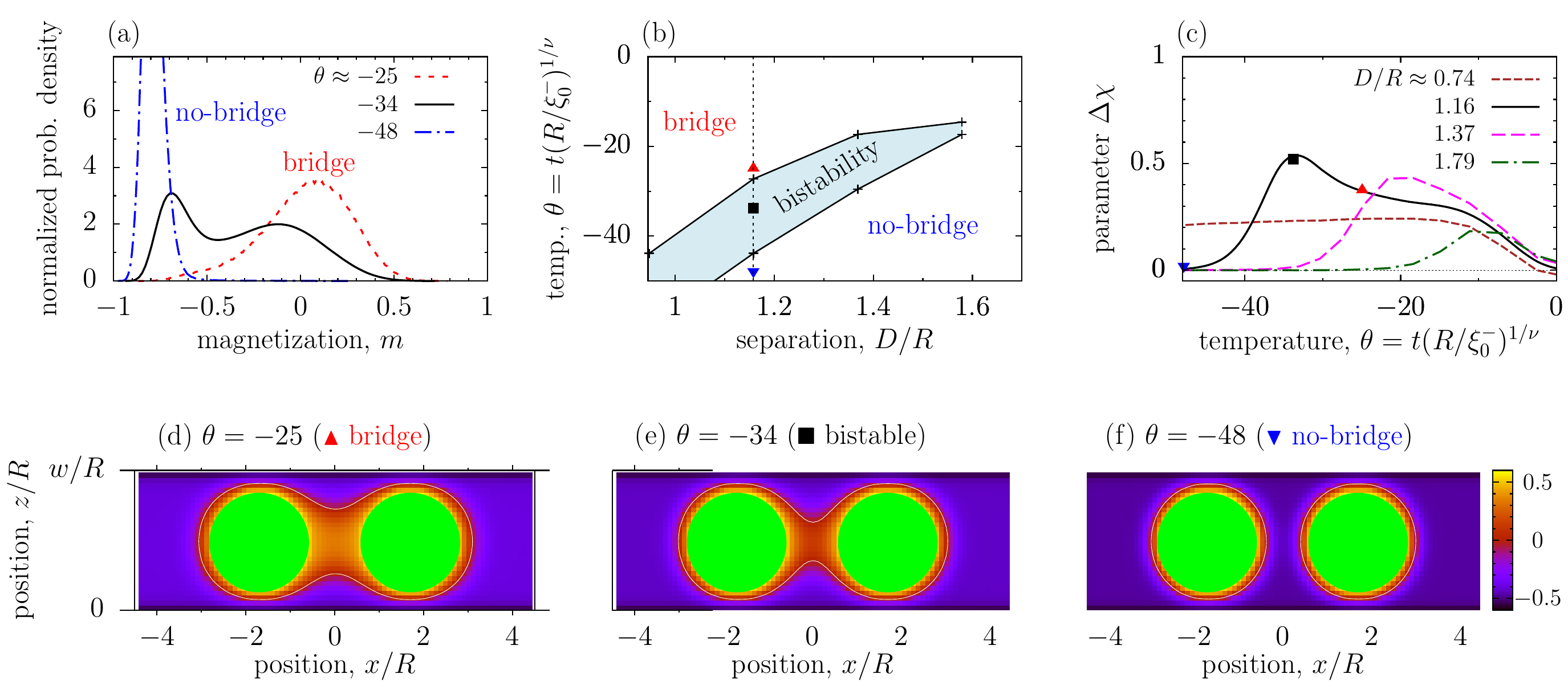}
    \caption{ \ftitle{Bridging transitions from Monte Carlo simulations of the Ising model.} (a) Probability distribution functions (PDFs), for three values of the rescaled temperature $\theta = - [R/\xi_-(t)]^{1/\nu} = t (R/\xi_0^-)^{1/\nu}$, corresponding to the bridge ($\theta \approx -25$) and the no-bridge ($\theta \approx -48$) configurations, and with both of them being stable or metastable ($\theta \approx -34$). The PDFs were calculated within a circle $C$ of radius $R_\mathrm{PDF}=9.5a$, laying in the plane perpendicular to the axis connecting the two colloids (\ie, in the plane $x=0$, see \fig{fig:mft}(a)). This axis crosses the center of the circle, which coincides with the midpoint between the two colloids. The scaled surface-to-surface distance between the colloids is $D/R=11/9.5 \approx 1.16$, where $R = 9.5a$ is the colloid radius and $a$ the lattice constant (see \sfig{fig:mc:R11.5} for $R=11.5a$). (b) Bridging phase diagram in the plane spanned by the rescaled temperature $\theta = t (R/\xi_0^-)^{1/\nu}$ and the scaled colloid separation $D/R$. The diagram shows the regions of the bridge and the no-bridge configurations separated by the two-phase coexistence domain (with linearly interpolated boundaries), in which both configurations are stable or metastable (compare \fig{fig:mft}(b)). The dotted vertical line and the symbols indicate the values of the colloid separation ($D/R \approx 1.16$) and of the rescaled temperatures used in (a) and (c). (c) Fluctuation parameter $\Delta \chi$ (\eq{eq:fpar}) as a function of the rescaled temperature $\theta$ for four values of the colloid-colloid separation $D/R$. The black curve in (c) corresponds to the dotted  vertical line in (b); accordingly the symbols in (b) and (c) correspond to each other, too. (d)-(f) ``Heatmaps'' of the average magnetization corresponding to the bridge, bistable, and no-bridge configurations, as indicated by the symbols in (b) and (c). The magnetization is expressed in units of the amplitude of the bulk magnetization below the critical point. In all plots the slit width is $w/R = 29/9.5 \approx 3.05$.
	\label{fig:mc}
}
\end{center}
\end{figure}

In order to study bridging transitions within MC simulations, we have computed the probability distribution function (PDF) of the normalized magnetization, 
\begin{align}
    m_C = \frac{1}{N_C} \sum_{i \textrm{ inside } C} \sigma_{i},
\end{align}
to have the value $m$. Here the sum runs over all lattice sites belonging to the interior of a planar contour line $C$; $N_C$ is the number of these sites. In the present study, the PDFs have been calculated within a circle with the same radius as the colloids, laying in the plane perpendicular to the axis connecting the colloids (\ie, in the plane $x=0$, see \fig{fig:mft}(a)). This axis vertically crosses the center of the circle, which coincides with the midpoint between the two colloids. The results are shown in \fig{fig:mc}(a) for three values of the rescaled  temperature. At low temperature ($\theta=-48$ in \fig{fig:mc}(a)), the PDF attains its maximum at a negative magnetization, consistent with the boundary conditions on the slit walls preferring the negative magnetization, indicating the \emph{absence} of a bridge (\fig{fig:mc}(f)). Closer to $T_c$ (\ie, $\theta = -25$ in \fig{fig:mc}(a)), the PDF is broader and the maximum is shifted towards positive magnetizations. This is due to a bridge forming between the colloids, which consists of positive spins, preferred by the colloid surfaces (\fig{fig:mc}(d)).

In the intermediate temperature regime, there are two peaks in the probability distribution, which are due to the bridge and no-bridge states; this corresponds to two metastable and stable branches obtained within MFT (\sfig{fig:mft:fe}, and, \cf, \fig{fig:force}(a)). The appearance or disappearance of a second peak in the PDF can thus be identified with the emergence or loss of metastability of the corresponding morphological phases. We have used these data in order to determine the  bridging `phase diagram'  shown in \fig{fig:mc}(b). As in MFT, the phase diagram consists of the regions of bridge and no-bridge configurations, which are separated by a bistability region, in which both states are stable or metastable. This two-phase coexistence  region shrinks as the colloid-colloid separation increases, and ends at a bridging critical point. (We could not precisely locate this latter critical point because -- due to the lattice nature of the Ising model -- the colloid-colloid separations are discrete.)

In order to gain more insight into the bridging transitions, we have computed the `fluctuation parameter', defined as
\begin{align}
\label{eq:fpar}
\Delta \chi = \chi (T, D) - \chi (T, D_\mathrm{max}),
\end{align}
where $D_\mathrm{max}$ is a sufficiently large separation, beyond which there is no bridge between the colloids (we took $D_\mathrm{max}$ as the maximum possible separation which can be implemented in our setup), and
\begin{align}
%\chi (T, D) = N_A \left(\langle \sigma^2\rangle - \langle \sigma\rangle^2 \right) / (k_BT)
    \chi (T, D) = \beta N_A \left(\langle \sigma^2\rangle - \langle \sigma\rangle^2 \right)
\end{align}
is the \emph{layer} susceptibility \cite{Binder-et:1995b}. Here $\sigma=(1/N_A)\sum_i \sigma_i$, with the sum running over all spins in the layer, $N_A = N_y\times N_z$ is the total number of spins in this layer, the brackets $\langle \cdots \rangle$ indicate the statistical average, and $\beta = J/(k_BT)$ is the inverse dimensionless temperature, as before. For the layer, we have chosen the plane perpendicular to the axis connecting the colloids and crossing it at the midpoint between the colloids (\ie, the plane $x=0$, see \fig{fig:mft}(a)). This parameter $\Delta \chi$ monitors excess intra-layer fluctuations due to bridge formation, as discussed below.

Close to criticality $\theta = 0$, the parameter $\Delta \chi$ increases upon decreasing $\theta$ for all colloid separations considered (\fig{fig:mc}(c)). This is the case because a bridge-like structure develops between the colloids, which increases the intra-layer fluctuations due to the formation of the interface between the positive (inside the bridge) and negative (outside of the bridge) magnetizations. At small separations ($D/R \approx 0.74$ in \fig{fig:mc}(c)), $\Delta \chi$ remains nearly constant as the temperature decreases further (\ie, it becomes more negative), because the bridge is and remains the only stable configuration. This can be contrasted with a larger separation ($D/R \approx 1.79$ in \fig{fig:mc}(c)), for which there is no bridge at sufficiently low temperatures, and hence $\Delta \chi$ decays to zero for further decreasing temperature. However, for intermediate distances ($D/R \approx 1.16$ and $1.37$ in \fig{fig:mc}(c)), $\Delta \chi$ exhibits pronounced peaks, which are due to bistability. In these cases, in addition to the contribution of the interface fluctuations, the peaks in $\Delta \chi $ emerge also because both the bridge and the no-bridge states contribute to the statistical average.

%%%%%%%%%%%%%%%%%%%%%%%%%%%%%%%%%%%%%%%%%%%%%%%%%%%%%%%%%%%%%%%%%%%
\subsection{Rounding off and sharpening of the bridging transition}

\begin{figure}[th]
\begin{center}
\includegraphics[width=\textwidth]{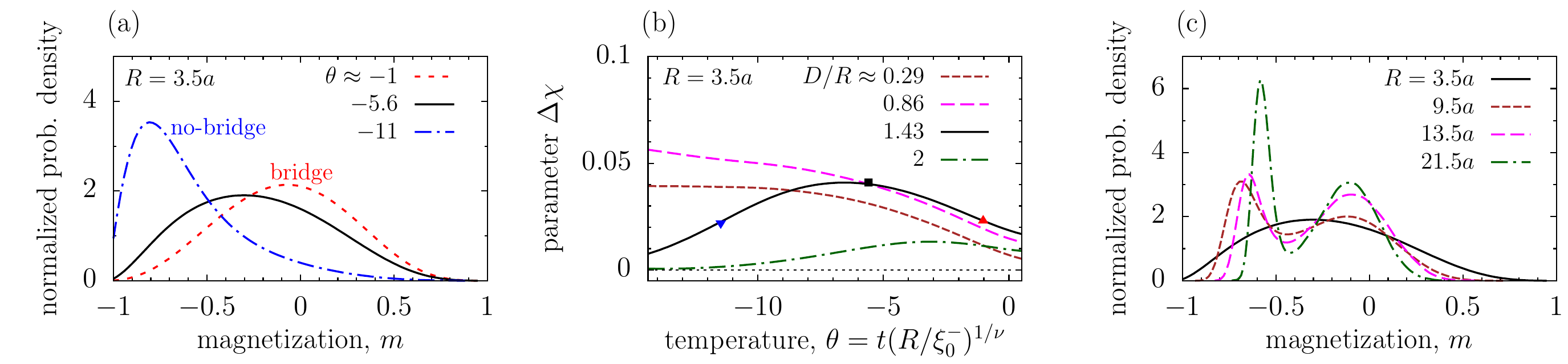}
    \caption{\ftitle{Absence of bistability for small colloids and sharpening of the bridging transitions for large colloids.} (a) Probability distribution functions (PDFs)  from Monte Carlo simulations for three values of the rescaled temperature $\theta = t (R/\xi_0^-)^{1/\nu}$  for colloids of radius $R=3.5a$, where $a$ is the lattice constant. The PDFs were calculated within a circle of radius $R_\mathrm{PDF}=R$, laying in the plane perpendicular to the axis connecting the two colloids, such that this axis crosses the center of the circle, which coincides with the midpoint between the colloids. The surface-to surface distance between the colloids is $D/R=5/3.5 \approx 1.43$. The bridge state continuously transforms into the no-bridge state without developing a two-peak structure, which would indicate two-phase coexistence (compare with (c) and \fig{fig:mc}(a)). (b) Fluctuation parameter $\Delta \chi$ (see \eq{eq:fpar}) as a function of the rescaled temperature $\theta$ for four values of the colloid-colloid separation $D/R$. The symbols denote the three values of the reduced temperature used in (a). There is no (second) peak in $\Delta \chi$ which would indicate the bridge--no-bridge bistability (compare \fig{fig:mc}(c)). In (a) and (b), the slit width is $w/R = 11/3.5 \approx 3.14$. (c) Probability distribution functions in the bistability region for four values of the colloid radius $R$, showing a sharpening of the transition upon increasing $R$. Note, however, that there is no bistability for $R =3.5a$. The parameters used here are: slit width $w/R = 11/3.5 \approx 3.14 $, colloid-colloid separation $D/R=5/3.5 \approx 1.43$, and rescaled temperature $\theta \approx -27$ ($R=3.5a$, see (a)); $w/R = 29/9.5 \approx 3.1 $, $D/R=11/9.5 \approx 1.16$, and $\theta \approx -34$ ($R=9.5a$, \fig{fig:mc}(a));  $w/R = 41/13.5 \approx 3.04$, $D/R=15/13.5 \approx 1.11$, and $\theta \approx -43$ ($R=13.5a$); and  $w/R = 65/21.5 \approx 3.02$, $D/R=23/21.5 \approx 1.07$, and $\theta \approx -60$ ($R=21.5a$).
	\label{fig:mc:nobridge}
}
\end{center}
\end{figure}

It is known that bridging transitions are rounded off by fluctuations \cite{Bauer-et:2000, Labbe-Laurent:17:jcp, privman_fisher:83:jsp}. In the Monte Carlo simulations discussed above, this is reflected by the fact that, unlike in MFT, we could not detect a sharp first-order transition between the bridge and the no-bridge states. Instead we observed a bistable region, in which both configurations were stable or metastable (\fig{fig:mc}(b)). The order parameter profile in such a bistability region is an average of both configurations, and appears as a state with or without a bridge, depending on the relative contribution of each of these two states (see \fig{fig:mc}(e) for the case in which the observed average configuration in the bistable region appears to be the one with a bridge).

In order to investigate the rounding off of the bridging transition in more detail, we have studied its dependence on the colloid radius. Interestingly, we observed no bistability for small colloids. \Fig{fig:mc:nobridge}(a) shows the normalized probability density functions (PDFs) for colloids of radius $R=3.5a$, which demonstrate that there is a continuous transformation between smeared-out bridge and no-bridge states, without a bistability region in between (\ie, two peaks in the PDF). A similar result has been reported in \myrefs{malijevsky:molphys:15,malijevsky15} for a bulk system (\ie, two colloids without a slit) with a square-well fluid-fluid interaction potential. In the present case, the absence of bistability shows up also in the fluctuation parameter $\Delta \chi$, which does not exhibit a strong peak associated with the presence of bistable bridge and no-bridge configurations (compare \fig{fig:mc:nobridge}(b) and \fig{fig:mc}(c)).

\Fig{fig:mc:nobridge}(c) shows the normalized PDFs in the \emph{bistable region} for a few values of the colloid radius $R$. This figure demonstrates that the maxima of the PDFs, corresponding to the bridge and no-bridge states, sharpen and become more pronounced, and the minimum between them deepens, upon increasing the colloid radius. Thus, in line with the arguments provided in \myref{Bauer-et:2000}, the rounding off of the bridging transition weakens for larger colloids. We therefore expect a \latin{de facto} sharp transition (accompanied by hysteresis) for experimentally relevant, macroscopically large colloids, similarly as predicted by MFT.

%%%%%%%%%%%%%%%%%%%%%%%%%%%%%%%%%%%%%%%%%%%%%%%%%%%%%%
\subsection{Capillary forces between confined colloids}

\begin{figure}[t]
\begin{center}
\includegraphics[width=0.8\textwidth]{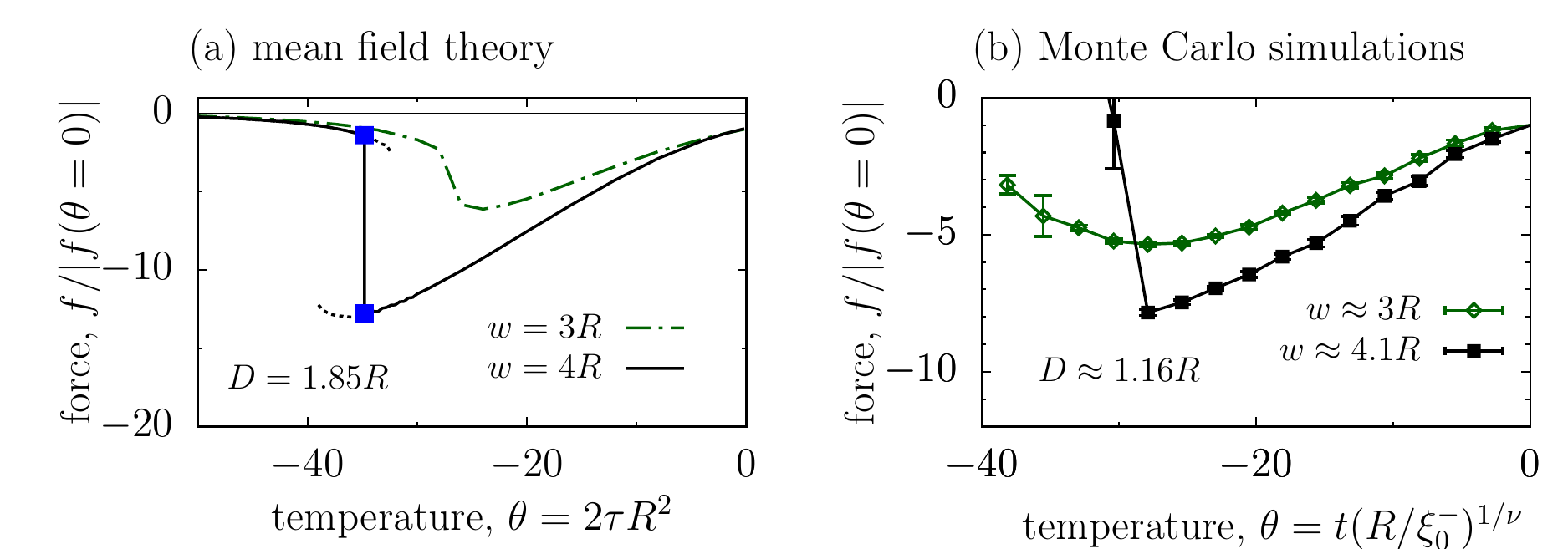}

    \caption{ \ftitle{Capillary forces between colloids confined to a slit filled with a two-phase fluid.} (a) Capillary force as a function of the rescaled temperature $\theta = - [R/\xi_-(t)]^{1/\nu}$ from mean field calculations for two slit widths $w$. Here $\xi_- = \xi_0^- (-t)^{-\nu}$ is the correlation length in the ordered phase, $\nu$ is the critical exponent, and $t = (T-T_c)/T_c$ is the reduced temperature relative to the bulk critical temperature $T_c$. Within mean field theory one has $\nu=1/2$ and $\theta = 2 \tau R^2$, where $\tau=t/(\sqrt{2} \xi_0^-)^2$. The surface-to-surface distance between the colloids is $D/R = 1.85$, where $R$ is the colloid radius. The short dotted lines show metastable branches for $w=4R$. No metastability has been found for $w=3R$. The blue squares denote the location of the first-order bridging transition (see \fig{fig:mft}(b)). (b) Capillary force as a function of the rescaled temperature $\theta = - [R/\xi_-(t)]^{1/\nu} = t (R/\xi_0^-)^{1/\nu}$ as obtained from Monte Carlo simulations for the colloid separation $D/R = 11/9.5 \approx 1.16$ and for two values of the slit width $w$. In both (a) and (b), the force is expressed in terms of the force at the critical point. A negative force means that the colloids attract each other. 
    \label{fig:force}
}
\end{center}
\end{figure}

    First, we calculated the force acting between the colloids by using mean field theory. For a wide slit ($w=4R$ in \fig{fig:force}(a)), and starting from bulk criticality $\theta=0$, the absolute value of the force increases drastically as the temperature decreases towards the bridging transition. Remarkably, the strength of the force attains values an order of magnitude stronger than the corresponding force at the bulk critical point of the fluid. At the transition, when the bridge breaks, the force drastically drops and virtually vanishes. The dotted lines in \fig{fig:force}(a) show the force associated with the metastable states, suggesting the possibility to observe hysteresis. In line with \myref{vasilyev:18:sm}, we found that the hysteresis loop widens upon decreasing the colloid-colloid separation, and the force can become even a few orders of magnitude stronger than at the bulk critical point (\sfig{fig:mft:force}) \cite{vasilyev:18:sm}. 

   For a narrower slit, when there is no first-order transition and the transformation between the bridge and no-bridge configurations proceeds continuously (see $w=3R$ in \fig{fig:force}(b)), accordingly the capillary force also becomes a continuous function of temperature (\fig{fig:force}(a)). Interestingly, the strength of the force becomes significantly reduced upon narrowing the slit. This reduction is likely due to the closer proximity of the slit walls, which impairs the bridge formation and hence weakens the force.

Our Monte Carlo simulations confirm that the strength of the capillary force decreases with decreasing the slit width (\fig{fig:force}(b)). For the wider slit, $w \approx 4.1R$, the force drops practically to zero at low temperatures (at $\theta \approx -9$ in \fig{fig:force}(b)), similarly as in MFT, which is likely due to the occurrence of a first-order bridging transition. It must be noted, however, that in this region our method for computing the force becomes inaccurate due to the onset of bistability (notice the wide error bar at $\theta \approx - 9$ for $w \approx 4.1 R$ in \fig{fig:force}(b)).

\section{Conclusion}
\label{sec:concl}

We have studied the formation and breaking of capillary bridges between colloids immersed in a two-phase fluid confined to a slit, with the slit walls and the colloid surfaces having opposite preferences for one of the fluid phases (\eg, for the species of a binary liquid mixture). We reported that, depending on the slit width and colloid-colloid separation, the formation and breaking of bridges can proceed via a first-order phase transition, or continuously without a \latin{bona fide} transition. 

By using mean field theory (MFT), we determined a bridging phase diagram. It consists of regions in which the bridge and the no-bridge states are stable; these regions are separated by a line of first-order phase transitions, which ends at a critical point. At small colloid-colloid separations, the transitions occur at low rescaled temperatures (\ie, far away from the bulk critical point of the fluid) and extend towards criticality as the separation increases (\fig{fig:mft}(b)). This phase behavior is similar to the case of colloids in the bulk (\ie, without the slit) \cite{Okamoto-et:2013, Labbe-Laurent:17:jcp}. In a slit, however, the bridging transitions are additionally influenced by the slit walls. In particular, we found that the bridging critical point is shifted towards smaller separations as the slit width decreases, so that a transformation between the bridge and no-bridge states can become continuous, without a transition (\fig{fig:mft}(b)).

We have calculated the capillary force acting between the colloids (\fig{fig:force}). In the vicinity of a first-order transition, concomitant by hysteresis, the force can vary significantly along the hysteresis loop. It is an order of magnitude stronger than the forces at fluid bulk criticality, provided there is a bridge between the colloids, but it drops abruptly when the bridge breaks. At sufficiently large colloid-colloid separations, the transition may disappear upon decreasing the slit width, and the force becomes a continuous function of temperature. Remarkably, in this case the strength of the forces weakens considerably (\fig{fig:force}). This change in the behavior of the force can be used as an experimental indicator of the occurrence of the bridging transition.

We have also performed Monte Carlo simulations of the same system by using the Ising model, which mimics an incompressible binary liquid mixture or a simple fluid. By calculating the probability density function for the magnetization, and identifying its maxima with the bridge and the no-bridge states, we confirmed the presence of wide bistability regions, in which these two states are stable or metastable. The corresponding phase diagram has a structure similar to that predicted by MFT (see \figss{fig:mc}(b) and \ref{fig:mft}(b), respectively). However, we found no bistability in the case of small colloids (such as for $R = 3.5a$, where $a$ is the lattice constant, see \figss{fig:mc:nobridge}(a) and (b)). This result is consistent with the findings reported in \myrefs{malijevsky:molphys:15,malijevsky15} (see also \myrefs{Bauer-et:2000, Labbe-Laurent:17:jcp, privman_fisher:83:jsp}), where it has been demonstrated that, for a bulk system (\ie, in the absence of the slit) with a square-well fluid-fluid interaction, the bridging transitions occur only if the colloids are sufficiently large compared to the microscopic length scale. We also showed that, upon increasing the colloid radius, the maxima of the probability density function become more pronounced, and a minimum between them deepens, which indicates a sharpening of the (first-order) bridging transition \cite{Bauer-et:2000, Labbe-Laurent:17:jcp, privman_fisher:83:jsp} (\fig{fig:mc:nobridge}(c)). It would be interesting to investigate experimentally such a transition sharpening, for instance by comparing the formation and breaking of capillary bridges for small and large colloids (such as quantum dots \cite{Marino2016, Marino2019} and micrometer-sized colloids \cite{Hertlein-et:2008, paladugu:15:nonaddcas}).

%\section*{Data Availability Statement}

%The data that support the findings of this study are available from the corresponding author upon reasonable request.

\bibliography{bridge}

\end{document}